\documentclass[a4paper,fleqn,usenatbib]{mnras}

\usepackage{graphicx}	
\usepackage{amsmath}	
\usepackage{mathtools}
\usepackage{amssymb}	
\usepackage{natbib}
\usepackage[flushleft]{threeparttable}
\usepackage{hyperref}

\usepackage[T1]{fontenc}
\usepackage{ae,aecompl}

\title[The IR-luminous progenitors of high-$z$ quasars]{The infrared-luminous progenitors of high-$z$ quasars}

\author[M. Ginolfi et al.]{
M. Ginolfi$^{1,2}$\thanks{E-mail: michele.ginolfi@inaf.it}, 
R. Schneider$^{2,3}$,
R. Valiante$^{1}$,
E. Pezzulli$^{2,3}$,
L. Graziani$^{2}$,\newauthor
~S. Fujimoto$^{4}$,
and R. Maiolino$^{5,6}$
\\\\
$^{1}$INAF/Osservatorio Astronomico di Roma, Via di Frascati 33, 00040 Monte Porzio Catone, Italy\\
$^{2}$Dipartimento di Fisica, Sapienza Universit\`{a} di Roma, Piazzale Aldo Moro 5, I-00185, Roma, Italy\\
$^{3}$INFN, Sezione di Roma I, P.le Aldo Moro 2, 00185 Roma, Italy\\
$^{4}$Institute for Cosmic Ray Research, The University of Tokyo, Kashiwa, Chiba 277-8582, Japan\\
$^{5}$Cavendish Laboratory, University of Cambridge, 19 J.J Thomson Ave., Cambridge CB3 0HE, UK\\
$^{6}$Kavli Institute for Cosmology, University of Cambridge, Madingley Road, Cambridge CB3 0HA, UK
}

\date{Accepted XXX. Received YYY; in original form ZZZ}

\pubyear{2018}

\begin{document}
\label{firstpage}
\pagerange{\pageref{firstpage}--\pageref{lastpage}}
\maketitle

\begin{abstract}
Here we explore the infrared (IR) properties of the progenitors of high-$z$ quasar host galaxies.
Adopting the cosmological, data constrained semi-analytic model \texttt{GAMETE/QSOdust}, we simulate several independent merger histories of a luminous quasar at $z\sim 6$, following black hole growth and baryonic evolution
in all its progenitor galaxies.
We find that a fraction of progenitor galaxies (about 0.4 objects per single luminous quasar) at $6.5 <z< 8$ has an IR luminosity of $L_{\rm{IR}}>10^{13} ~L_{\odot}$ (hyper-luminous IR galaxies; HyLIRGs).
HyLIRGs progenitors reside in the most massive halos, with dark matter (DM) masses of $M_{\rm{DM}} \sim 10^{12.5}-10^{13} ~ M_\odot$.
These systems can be easily observed in their $\sim1$ mm-continuum emission in a few seconds of  integration time with the Atacama Large Millimeter/submillimeter Array (ALMA),
and at least 40\% of them host nuclear BH activity that is potentially observable in the soft and hard X-ray band. 
Our findings are in line with recent observations of exceptional massive DM halos hosting HyLIRGs at $z\sim7$, suggesting that $z\sim 6$ luminous quasars are indeed the signposts of these observed rare peaks in the high-$z$ cosmic density field, and that massive IR-luminous galaxies at higher $z$ are their natural ancestors. 
\end{abstract}

\begin{keywords}
Galaxies: evolution, high-redshift; quasars: general, supermassive black holes; infrared: galaxies; submillimetre: galaxies
\end{keywords}



\section{Introduction}

Observations of dusty star-forming galaxies (DSFGs) at high-redshift, often selected at (sub-)millimeter wavelengths (submillimeter galaxies, or SMGs; \citealp{Smail1997,Hughes1998,Chapman2005,Hayward2013}), can provide unique insights into our understanding of the early formation of massive galaxies (see e.g., \citealp{Blain2002,Casey2014} for reviews).
Because of their large dust content, DSFGs emit most of their luminosity at infrared (IR) wavelengths ($8-1000 ~\rm{\mu m}$) and are typically considered ultra-luminous IR galaxies (ULIRGs), reaching IR luminosities ($L_{\rm{IR}}$) of $L_{\rm{IR}} >10^{12} ~L_{\odot}$ (\citealp{Kovacs2006,Coppin2008,Hayward2011}).
Spectroscopic surveys targeting the redshift distributions of SMGs indicate that the bulk of the DSFGs population peaks at $z\sim 2-4$ (e.g., \citealp{Chapman2005,Strandet2016}), encompassing the peaks of supermassive black hole (SMBH) accretion (\citealp{Cattaneo2003,Hopkins2007}) and cosmic star formation activity (\citealp{Madau2014}).
However, a significant tail of higher redshift DSFGs appears to be already in place at $z > 5$ (e.g., \citealp{Riechers2013,Riechers2017,Walter2012,Weiss2013,Strandet2017}). 
The latter are often found to be extreme hyper-luminous IR galaxies (HyLIRGs), reaching infrared luminosities of $L_{\rm{IR}} >10^{13} ~L_{\odot}$ and star formation rates (SFRs) exceeding $1000 ~ M_{\odot}~\rm{yr}^{-1}$, likely tracing the result of strong dynamical interactions (i.e., major mergers) and intense gas accretion events occurring in the densest regions of the early Universe (e.g., \citealp{Riechers2011,Riechers2017,Ivison2011,Capak2011,Oteo2016,Pavesi2018}).
\newline
Recently, \cite{Marrone2018} reported high-resolution Atacama Large Millimeter/submillimeter Array (ALMA) observations of SPT0311-58, a IR-luminous system at $z = 6.9$, originally identified in the 2500 deg$^2$ South Pole Telescope (SPT) survey (\citealp{Carlstrom2011}).
ALMA reveals this source to be a pair of interacting extremely massive and IR-luminous star-bursting galaxies: 
SPT0311-58W, a HyLIRG with $L_{\rm{IR}}  = 3.3 \pm 0.7 \times 10^{13} ~L_{\odot}$ and a $\rm{SFR} \sim 2900 ~M_\odot~\rm{yr}^{-1}$, and 
SPT0311-58E, a ULIRG with $L_{\rm{IR}} = 4.6 \pm 1.2 \times 10^{12} ~L_{\odot}$ and $\rm{SFR} \sim 540 ~M_\odot~\rm{yr}^{-1}$.
Using different proxies, \cite{Marrone2018} estimated the DM halo mass hosting this system to be $M_{\rm{DM}} = (1.4 - 7) \times 10^{12} ~M_\odot$, showing that SPT0311-58 marks a rare peak
in the cosmic density field at this early cosmic time and it lies close to the exclusion curve predicted by the current structure formation paradigm. 
\newline
\newline
Many models of SMBH-galaxy co-evolution suggest an evolutionary link between ULIRGs and quasars (quasars; \citealp{Sanders1988a,Sanders1988b,Silk1998,Springel2005,DiMatteo2005,Hopkins2006,Hopkins2008}, and \citealp{Alexander2012} for a review).
ULIRGs may represent the initial, heavily obscured, stages of quasars, which, after shedding the surrounding dust through energetic galaxy-scale outflows (e.g., \citealp{Maiolino2012, Harrison2012,Fan2018}), evolve into a ultraviolet (UV)/optical bright unobscured phase.
\newline
In this work, we explore 
the hierarchical merger histories of $z\sim 6$ quasars to investigate their connection with ULIRGs and HyLIRGs. 
To this aim, we use \texttt{GAMETE/QSOdust} (hereafter \texttt{GQd}, \citealp{Valiante2014,Valiante2016}), a semi-analytic model tested to reproduce the observed properties of a sample of high-$z$ quasars (\citealp{Valiante2014}).
\texttt{GQd} consistently follows the formation and evolution of nuclear black holes (BHs) and their host galaxies, accounting for star formation, interstellar medium (ISM) chemical evolution (metals and dust) and mechanical feedback, along different merger histories of the parent DM halo.
Therefore, \texttt{GQd} is a suitable tool to (i) investigate the physical properties of the ancestors at $z\sim7 - 9$ of luminous quasars at $z\sim 6$; (ii) statistically quantify the number density of expected ULIRG and HyLIRG progenitors, and (iii) probe whether $z\sim6$ luminous quasars can be the signposts 
of the rare peaks in the cosmic density field observed at $z\sim 7$ (i.e., \citealp{Marrone2018}).
\newline
The paper is organized as follows. 
In Section \ref{sec: model}, we briefly describe the \texttt{GQd} model.
Section \ref{sec: results} and \ref{sec: conclusions} show the results of this work and a discussion of their implications.
%
Throughout the paper, we assume a $\Lambda$CDM cosmology with  $\Omega_{\rm{m}} = 0.24$, $\Omega_{\rm{\Lambda}} = 0.76$, $\Omega_{\rm{b}} = 0.04$, and $H_0 = 73 ~\rm{km ~ s^{-1} ~ Mpc^{-1}}$.

\section{Model Description}\label{sec: model}

\texttt{GQd}\footnote{The model \texttt{GAMETE} (GAlaxy MErger Tree and Evolution) was originally developed to study the formation and cosmological evolution of local, Milky-Way like galaxies (\citealp{Salvadori2007,Salvadori2008}).} is a semi-analytic, data-constrained model aimed at studying the formation and evolution of high-redshift quasars and their host galaxies in a cosmological framework.
In this section we summarize the main features of \texttt{GAMETE/QSOdust}, referring the reader to \cite{Valiante2011,Valiante2014,Valiante2016} for a thorough description of the model. 
\newline
\texttt{GQd} has been tested to reproduce the observed properties of quasars at $z>5$, such as the BH mass, the molecular gas and dust masses in the ISM (\citealp{Valiante2014}). 
Following \cite{Valiante2016}, in this work we focus on the evolution of a SDSS J1148+5251 (hereafter J1148)-like quasar at $z=6.4$, one the best studied high-$z$ luminous quasars%
\footnote{Absolute AB magnitude of the continuum in the rest-frame at 1450 \AA ~of $M_{1450} = −27.82$ (\citealp{Fan2003}).}
(e.g., \citealp{Willott2003,Walter2004,Maiolino2005,Cicone2015}), and we use it as a prototype for the general class of luminous $z \sim 6$ quasars.

\begin{table}
	\caption{Observed and inferred properties of the quasar SDSS J1148+5251 at $z=6.42$.
		}
	\label{tab:properties_quasar}
		\begin{tabular}{ccccc} 
			\hline
			$^{(a)}L_{\rm{FIR}}$&
			$^{(b)}L_{\rm{bol}}$& 
			$^{(a)}$SFR&
			$^{(a)}$$M_{\rm{dust}}$&
			$^{(c)}$$M_{\rm{BH}}$\\

			[$10^{13} L_{\odot}$]&
			[$10^{14} L_{\odot}$]&
			[$10^3 M_\odot \rm{yr}^{-1}$]&
			[$10^8 M_\odot$]&
			[$10^9 M_\odot$]\\ 
			\hline 
	
			$2.2\pm0.33$ &
			$1.36\pm0.74$ &
			$2.0\pm0.5$&
			$3.4^{+1.38}_{-1.54}$&
			$4.9\pm2.5$	\\ 
			\hline
		\end{tabular}
				\begin{tablenotes}
				\item (a) The values of $L_{\rm{FIR}}$, SFR, and $M_{\rm{dust}}$ have been computed by 
				\cite{Valiante2011, Valiante2014}.
				\item (b) The bolometric luminosity, $L_{\rm{bol}}$, is estimated from the observed flux at 1450 \AA~ (\citealp{Fan2003}) using the bolometric correction by \cite{Richards2006}.
				\item (c) The black hole mass, $M_{\rm{BH}}$, is estimated from the Mg$_{\rm{II}}$ doublet and the $\lambda = 3000 ~$\AA~ continuum (\citealp{DeRosa2011}). 
				\end{tablenotes}
\end{table}

\subsection*{The hierarchical merger history}  

We first reconstruct different hierarchical merger histories (merger trees) of a $M_{\rm{DM}} = 10^{13} M_\odot$ DM halo hosting a J1148-like quasar at $z=6.4$, using  a binary Monte Carlo algorithm accounting for mass infall (e.g., \citealp{Volonteri2003}), based on the Extended Press-Schechter theory (e.g., \citealp{Lacey1993}).
This approach enables us to produce a large (statistically meaningful) number of random, semi-analytical merger trees of the same DM halo, simultaneously resolving the first collapsed objects.
The DM resolution mass, $M_{\rm res}$, is chosen to resolve the first collapsed objects where the gas is able to cool and form stars, i.e., the so-called \textit{mini-haloes}, with masses of $M_{\rm h} \sim 10^{6} M_\odot$ and virial temperatures of $T_{\rm vir}< 10^4 ~\rm{K}$ (see \citealp{Bromm2013} for a review).
Given this requirement, we simulate the merger trees adopting a $M_{res}$ of:
\begin{equation}
M_{res}(z_i) = 10^{-3} ~ M_{h}(z_0) ~ \left( \dfrac{1+z_i}{1+z_0}  \right)^{-7.5},
\end{equation}
where $M_{h}(z_0)$ is the host DM halo at redshift $z_0 = 6.4$.
Mini-haloes dominate the halo mass spectrum at very high-$z$; at $z\lesssim14$ the merger tree halo population is dominated instead by Ly$\alpha$-cooling halos, namely DM halos with $T_{\rm vir} \geq 10^4 \rm{K}$ (see \citealp{Valiante2016} for details).
\newline 
By construction, the entire population of halos along the merger trees, that we call \textit{progenitors}, contribute to assembling the simulated quasar host galaxies. 
However, when specified within the text, we will restrict the analysis to massive progenitors, hosted by DM halos with $M_{\rm{DM}}>10^{11} M_{\odot}$. 
This mass-selection, as discussed in Sec. \ref{sec: results}, reflects our interest of studying the most IR-luminous galaxies in the simulation.

\subsection*{BH growth and baryonic evolution}

Merger trees are used as an input to reconstruct the formation and evolution of the J1148-like quasar and its host galaxy through cosmic time.
The baryonic evolution inside each progenitor galaxy is regulated by processes of star formation, BH growth and feedback, and followed consistently through the cosmic mass assembly.
\newline
\newline
BHs growth is driven by:
\newline
\textit{- mergers:} we assume that in major mergers\footnote{A major merger occurs when two DM halos with mass ratio (less massive over the most massive) $\mu>1/4$ coalesce.}  the 
BHs present in the nuclei of the two interacting galaxies coalesce, forming a new more massive BH.
In minor mergers, the merger time-scale of the two BHs is of the order of the Hubble time or longer (see e.g., \citealp{Tanaka2009}). The less massive BH of the merging pair is assumed to remain 
as a satellite and we do not follow its evolution.
\newline
\textit{- gas accretion:} the accretion rate is described by a modified\footnote{We re-scaled by a factor $\alpha_{\rm BH} = 50$, 
to account for the higher central densities around accreting BHs (e.g., \citealp{DiMatteo2005}).} Bondi-Hoyle-Lyttleton (BHL) formula, capped at the Eddington rate (see \citealp{Valiante2014} for details).
We find that, while BH growth is mostly driven by mergers at very high-$z$ (i.e., $z \gtrsim 11$), gas accretion is the dominant growth mode at lower redshifts (\citealp{Valiante2016}).
\newline
\newline
At each snapshot of the merger tree, stars in the progenitor galaxies form according to a SFR proportional to the available gas mass, with an efficiency that is enhanced during major mergers (see \citealp{Valiante2014}).
Following star formation, supernovae (SNe) and asymptotic giant branch stars (AGB) progressively enrich the ISM of each galaxy with metals and dust according to their specific mass and metallicity-dependent yields and to their evolutionary time-scales.
We follow the life-cycle of metals and dust including physical prescriptions for dust processing in a two-phase ISM: SN shocks can destroy dust grains in the hot, diffuse medium while dust grains can grow in mass by accreting gas-phase heavy elements in the cold medium (see \citealp{Valiante2014} and \citealp{deBennassuti2014}, for details).
A fraction of the energy released by SN explosions (0.2\%) and BH accretion (0.25\%) is converted into kinetic energy of the gas in the host galaxy, driving gas outflows in the form of winds. 
While these efficiencies are lower than what generally adopted in hydrodynamical simulations\footnote{In one of the most popular descriptions of AGN-driven winds in hydrodynamical simulations,  $5\%$ of the
radiation energy is thermally coupled to the surrounding gas \citep{DiMatteo2005}. This difference may be due to the complex interaction of the outflow with the inhomogeneous ISM and the effects of radiative 
losses that can not be captured by our simple semi-analytical model.}
the predicted strength of mechanical feedback is in good agreement with observations of the outflowing gas in J1148 (\citealp{Maiolino2005,Maiolino2012,Valiante2012,Cicone2015}; see also \citealp{Bischetti2018}, who carry out a stacking analysis of a sample of 48 quasars at $4.5 < z < 7.1$ detected by ALMA in the [CII] 158 $\mu$m line, and find an outflow kinetic power that is $\sim 0.1\%$ of the AGN luminosity).

\section{THE IR PROPERTIES OF MASSIVE PROGENITORS OF HIGH-Z quasar HOST GALAXIES}\label{sec: results}

We calculate the IR luminosity of each galaxy by assuming that in the Rayleigh-Jeans part of the spectrum (in the limit of small frequencies, that is $h\nu \ll k_{\mathrm {B} }T$), dust radiates as a \textquoteleft grey-body\textquoteright~with an opacity coefficient per unit dust mass, $k_{\rm{d}}(\nu) = k_0~ (\nu/\nu_0)^{\beta}$, where $\nu_0$, $k_0$, $\beta$ depend on the adopted model of dust.
Integrating over the IR wavelength range $8-1000 ~\mu \rm{m}$  ($300 - 38000$ GHz),
we compute the corresponding $L_{\rm{IR}}$ as,
\begin{equation}
L_{\rm{IR}} = 4 \pi M_{\rm{dust}} \int k_{\rm{d}}(\nu) ~ B(\nu, T_{\rm{dust}}) ~d\nu,
\end{equation}
where $B(\nu, T_{\rm{dust}})$ is the Planck function for a dust temperature $T_{\rm{dust}}$.
We estimate $L_{\rm{IR}}$, averaging over 7 different models of dust composition%
\footnote{Following \cite{Valiante2011} we use models of dust composition from \cite{Bertoldi2003}, \cite{Robson2004}, \cite{Beelen2006}, \cite{Weingartner2001}, \cite{Bianchi2007}.
In addition, we also consider the more recent \textquoteleft AC\textquoteright-model by \cite{Galliano2011} and \textquoteleft THEMIS\textquoteright~ by \cite{Jones2017} (see the review by \citealp{Galliano2017} for a discussion).}
(each of them providing a unique combination of the parameters $k_0, \nu_0, \beta$; see Table \ref{tab:dust_composition}) and we vary the dust temperature%
\footnote{Warm interstellar dust, associated to starburst regions, dominates the emission in the rest-frame far-IR (\citealp{Dunne2000, Wang2008, Valiante2016}).} 
in the range $T_{\rm{dust}} = [35 - 55] ~\rm{K}$.
\begin{table}
	\caption{Compilation of the $k_0$, $\lambda_0$ and $\beta$ parameters, defining the opacity coefficient per unit dust mass in the different models of dust composition used in this work.
	}
	\label{tab:dust_composition}
	\centering
	\begin{tabular}{cccc} 
		\hline
		Model&
		$k_0$& 
		$\lambda_0$&
		$\beta$\\
		
		(reference)&
		[cm$^2$/gr]&
		[$\mu$m]&
		[$\beta$]\\ 
		\hline 
		
		\cite{Bertoldi2003} &
		$7.5$ &
		$230$&
		$1.5$\\ 
		
		\cite{Robson2004} &
		$30$ &
		$125$&
		$2.0$\\ 
		
		\cite{Beelen2006} &
		$0.4$ &
		$1200$&
		$1.6$\\

		\cite{Weingartner2001}$^{(a)}$ &
		$34.7$ &
		$100$&
		$2.2$\\ 
		
		\cite{Bianchi2007} &
		$40$ &
		$100$&
		$1.4$\\ 
		
		THEMIS - \cite{Jones2017} &
		$6.4$ &
		$250$&
		$1.79$\\ 
		
		AC - \cite{Galliano2011} &
		$16$ &
		$160$&
		$1.7$\\ 
		\hline
	\end{tabular}
	\begin{tablenotes}
		\item (a) Fit to the model for the Small Magellanic Cloud in the spectral range $[40-200]$ $\mu$m.
	\end{tablenotes}
\end{table}
\begin{figure}
	\centering
	\includegraphics[width=1\columnwidth]{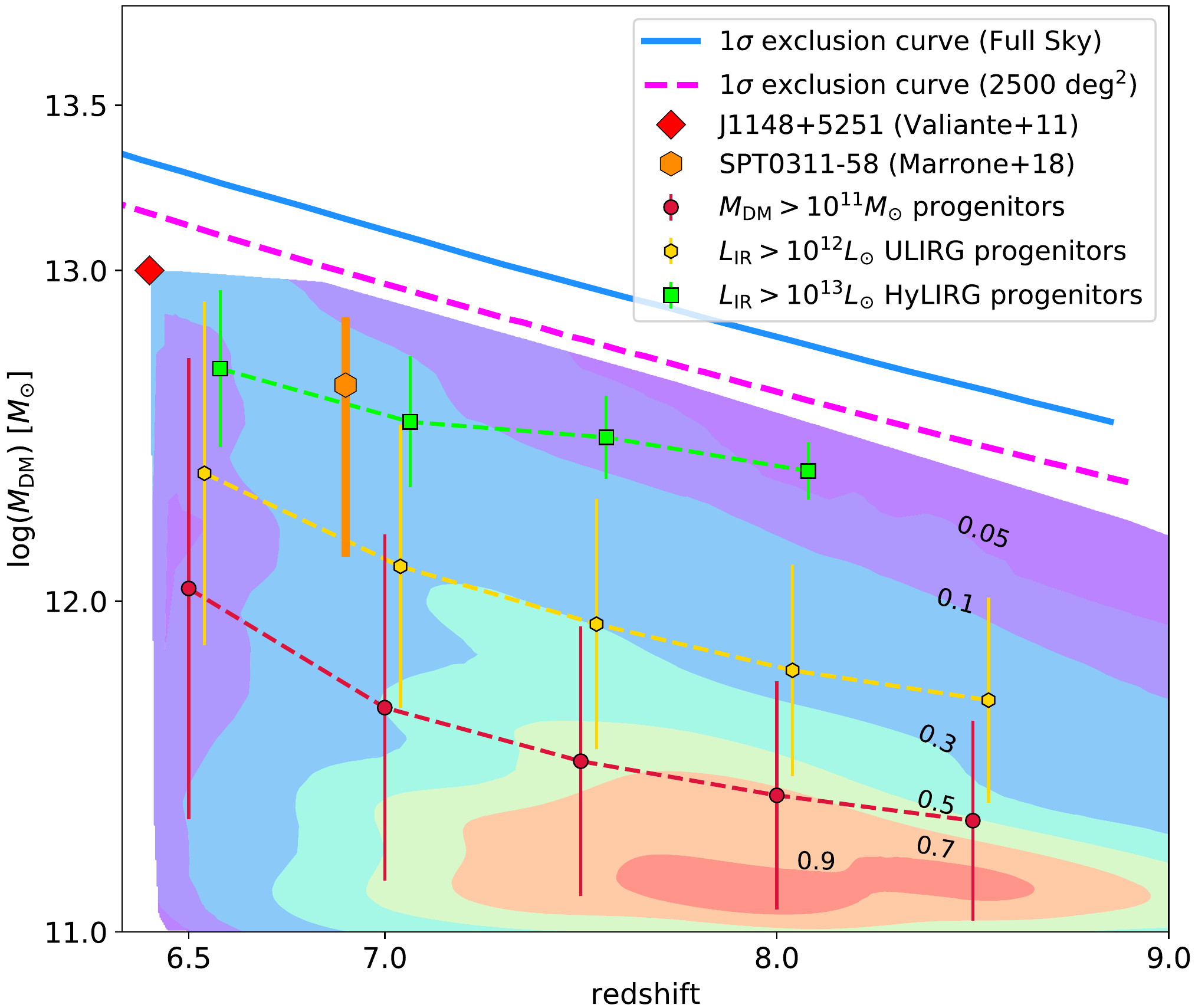}
	\caption{
		Density plot showing the $M_{\rm{DM}}$ distribution of DM halos hosting massive progenitor galaxies ($M_{\rm{DM}} > 10^{11} M_{\odot}$) of a J1148-like quasar host galaxy (red point), in the redshift interval $6.4 < z < 9$. 
		The density distribution is based on the total statistics of 10 different merger trees (see Sec. \ref{sec: model}); 
		the percentile levels are marked by the filled colour contours and labeled in the figure.
		The red dashed line and points represent the redshift evolution of the averaged $M_{\rm{DM}}$ of all massive progenitors ($M_{\rm{DM}} > 10^{11} M_{\odot}$).
		The gold and green lines and points represent the redshift evolution of the 
		average $M_{\rm{DM}}$ of halos hosting massive progenitors with $L_{\rm{IR}} > 10^{12} L_\odot$ (ULIRGs) and $L_{\rm{IR}} > 10^{13} L_\odot$ (HyLIRGs), respectively.
		Error-bars are indicative of the $1 \sigma$ standard deviation calculated in equispaced redshift bins (the slight offset of the bins position for red, gold and green points has been introduced to avoid overlapping and help visualization).	
		The orange point represents the DM halo mass hosting the HyLIRG system SPT0311-58, at $z=6.9$ (\citealp{Marrone2018}). 
		The blue solid and magenta dashed lines represent the 1$\sigma$  exclusion curves, i.e., the most massive halos that are expected to be observable within the whole sky and a 2500 deg$^2$ area (the size of the SPT survey in which SPT0311-58 was identified) respectively, as a function of redshift (\citealp{Harrison2013,Marrone2018}).
	}
	\label{fig:DM_analysis}
\end{figure}

\begin{figure}
	\centering
	\includegraphics[width=1\columnwidth]{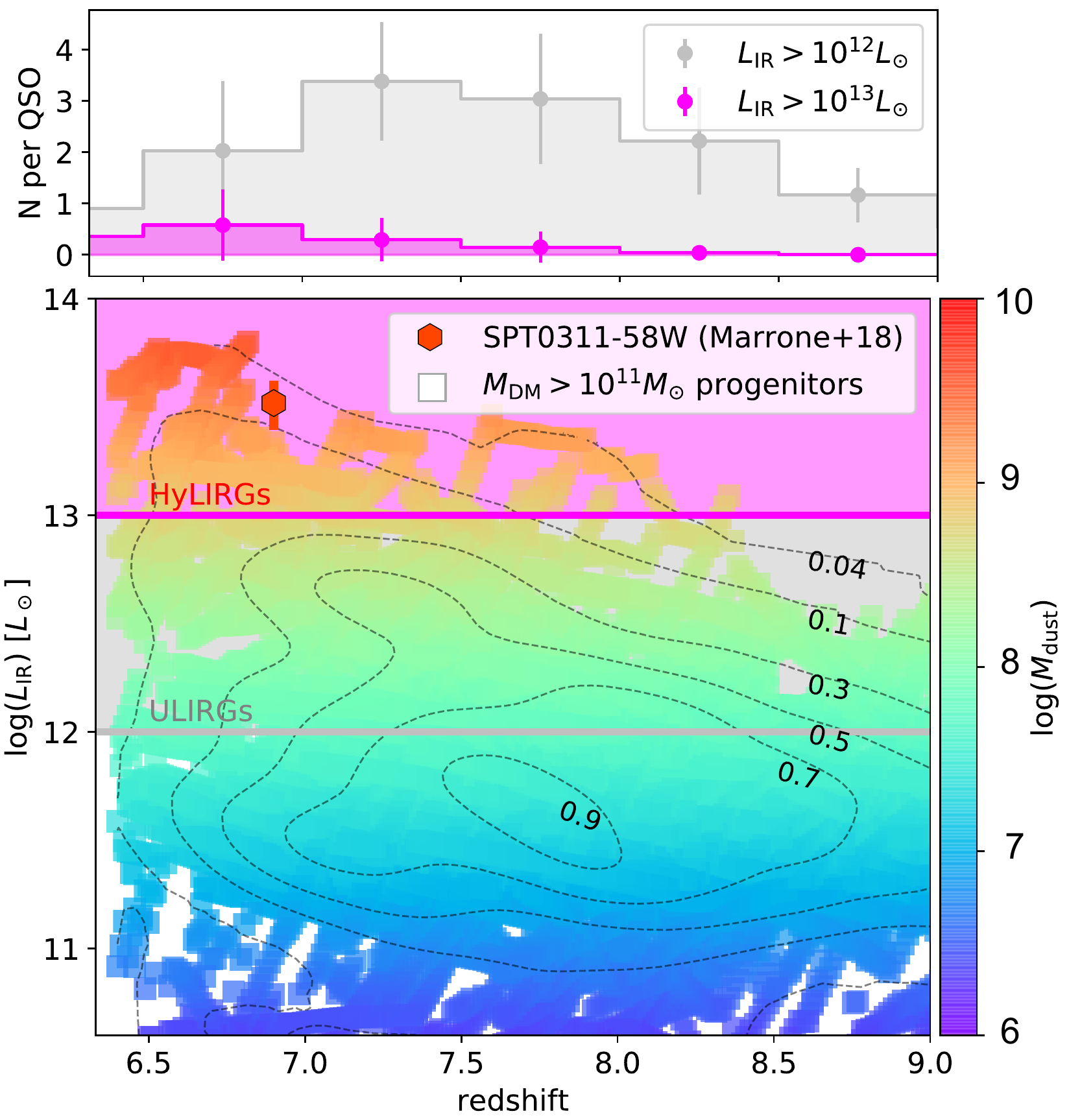}
	\caption{
		\textit{Bottom panel:} 
		redshift distribution of the $L_{\rm{IR}}$ in massive progenitors ($M_{\rm{DM}} > 10^{11} M_{\odot}$) of a J1148-like quasar host galaxy, in the redshift interval $6.4 < z < 9$, using the total statistics of 10 independent merger trees.
		The distribution is colour coded depending on the galaxy ISM dust mass.
		The black dashed lines represent the density contours with the corresponding percentiles labeled in the figure.
		The grey (magenta) filled regions define the $L_{\rm{IR}}$-range of ULIRGs (HyLIRGs).
		The red point represents the $L_{\rm{IR}}$ of SPT0311-58W, the most massive galaxy in the interacting system SPT0311-58 (\citealp{Marrone2018}).
		\textit{Top panel:} 
		average number of ULIRGs (grey points) and HyLIRGs (magenta points) at $6.4 < z < 9$, expected for each luminous quasar at $z\sim 6$. 
		Error-bars are indicative of the $1~\sigma$ standard deviation calculated over 10 different merger trees.
	}
	\label{fig:L_IR_analysis}
\end{figure}

Fig. \ref{fig:DM_analysis} and Fig. \ref{fig:L_IR_analysis} show respectively (i) the redshift distribution of simulated DM halo masses and (ii) the IR luminosities of massive progenitors ($M_{\rm{DM}} > 10^{11} M_{\odot}$) of J1148-like quasar host galaxies at $z\sim 6.4$, in all the simulated merger trees.
Fig. \ref{fig:DM_analysis} shows that, although the bulk of the mass distribution of massive progenitors is concentrated in the mass range $11 < \rm{log} (M_{\rm{DM}}/M_{\odot}) < 12$ 
(especially at $z\sim 8$; see percentile levels in the density plot), 
IR-luminous galaxies with $L_{\rm{IR}} > 10^{12} L_\odot$ occupy DM halos with $M_{\rm{DM}} > 10^{12} M_\odot$, at least in the redshift interval $6.4 < z < 7.5$.
In particular, we find that galaxies with HyLIRGs-like luminosity ($L_{\rm{IR}} > 10^{13} L_\odot$, consistent with the IR luminosity measured in SPT0311-58W; \citealp{Marrone2018}) 
reside in the most massive halos, with masses of about $M_{\rm{DM}}\sim10^{12.5}-10^{13} M_\odot$, consistent with the dynamical mass inferred for the interacting system SPT0311-58 (see orange hexagon in Fig. \ref{fig:DM_analysis}).
\newline
\newline
These findings indicate that exceptionally massive DM halos hosting HyLIRGs, as the $\sim 10^{12.6} M_\odot$ halo hosting SPT0311-58 at $z\sim 7$, can be found within the family tree of UV/optical bright quasars at $z\sim 6$.
\newline
In the next paragraphs we explore the expected physical properties of the IR-luminous progenitors of high-$z$ luminous quasar host galaxies, as well as their frequency (i.e., the expected number per quasar) and their observability in the mm and X-ray bands.

\subsection{How many IR-luminous progenitor galaxies do we expect for each $z\sim6$ luminous quasar?}
\label{sec: spatial_density}

In the top panel of Fig. \ref{fig:L_IR_analysis} we show the predicted number of IR-luminous progenitor galaxies in the redshift interval $6.4 < z < 9$ expected for each luminous J1148-like quasar at $z \sim 6.4$.
Each high-$z$ luminous quasar has, on average, about 2.8 ULIRG progenitors with $L_{\rm{IR}}>10^{12} L_{\odot}$, and about 0.4 HyLIRG progenitors with $L_{\rm{IR}}>10^{13} L_{\odot}$, between $z\sim 6-8$.
Recently, \cite{Jiang2016} using a complete sample of 52 quasars at $z>5.7$ selected from the  Sloan Digital Sky Survey (SDSS),  spanning a wide luminosity range of $-29.0 \leqslant M_{1450}  \leqslant -24.5$, found that the number density of bright quasars with $M_{1450}<-26.7$ at $z\sim6$ is $(0.4 \pm 0.1) ~\rm{Gpc}^{-3}$
\footnote{This value is consistent with the estimates provided by previous works, e.g.,
\cite{Fan2004} and \cite{Willott2010}.}. 
Combining this information with our estimate of the number of expected IR-luminous progenitor galaxies per quasar, we predict the spatial number density of HyLIRG (ULIRG) progenitors to be about
$0.16 \pm 0.08$ ($1.2 \pm 0.6$) $\rm{Gpc}^{-3}$ in the redshift range $z\sim 6-8$.
Since these values refer to the population of ULIRGs/HyLIRGs found within the merger trees of UV/optical bright quasars at $z \sim 6$, they provide lower limits on the effective number density of the overall population of IR-luminous galaxies at $z>6$.
Potential deviations might be also driven by the effect of the duty cycle on the number counting of high-$z$ quasars, resulting in a systematic under-estimation of their number densities in large single-epoch imaging surveys (see e.g., \citealp{RomanoDiaz2011}).
An alternative way to estimate the number density of the overall population of IR-luminous galaxies at $z>6$ comes from our finding that, at $6.5<z<8$, the most massive DM halos ($M_{\rm{DM}}>10^{12.5} M_{\odot}$) in the merger trees of simulated quasars are most likely to host HyLIRGs (see Fig. \ref{fig:DM_analysis}).
Assuming that this finding can be extended to all DM halos in the same mass regime (i.e., also halos that are not expected to evolve in a quasar at $z\sim6$), then the number density of the HyLIRGs population should be consistent with the number density of DM halos with $M_{\rm{DM}}>10^{12.5} M_{\odot}$ predicted by the $\Lambda$CDM cosmology\footnote{The $\Lambda$CDM-predicted number density of DM halos was obtained using \texttt{HMF\textit{calc}}, publicly available at \hyperlink{http://hmf.icrar.org/}{http://hmf.icrar.org/} (see \citealp{Murray2013}).} at $6.5<z<8$, i.e., about $2~\rm{Gpc}^{-3}$.

Fig. \ref{fig:L_IR_analysis} also shows that HyLIRGs are characterized by dust masses $\gtrsim 10^{8.7} M_\odot$ already at $z \sim 8 - 8.5$, while ULIRGs have dust masses in the range $10^{7.5} < M_{\rm{dust}}/M_\odot < 10^9$. 
\newline
Despite their large stellar masses, $\gtrsim 70\%$ of the dust mass has formed in the ISM of these galaxies through processes involving some form of grain growth%
\footnote{We note that the efficiency and even the physical nature of grain growth is still an active subject of study (e.g., \citealp{Zhukovska2016, Ceccarelli2018, Ginolfi2018}).}
(see \citealp{Valiante2011}, where the maximum dust masses obtained from stellar sources, e.g., SNe and AGB stars, is calculated as a function of $M_{\rm{star}}$, under the conservative assumption that dust grains are not destroyed by interstellar shocks).
\begin{figure*}
	\centering
	\includegraphics[width=0.9\textwidth]{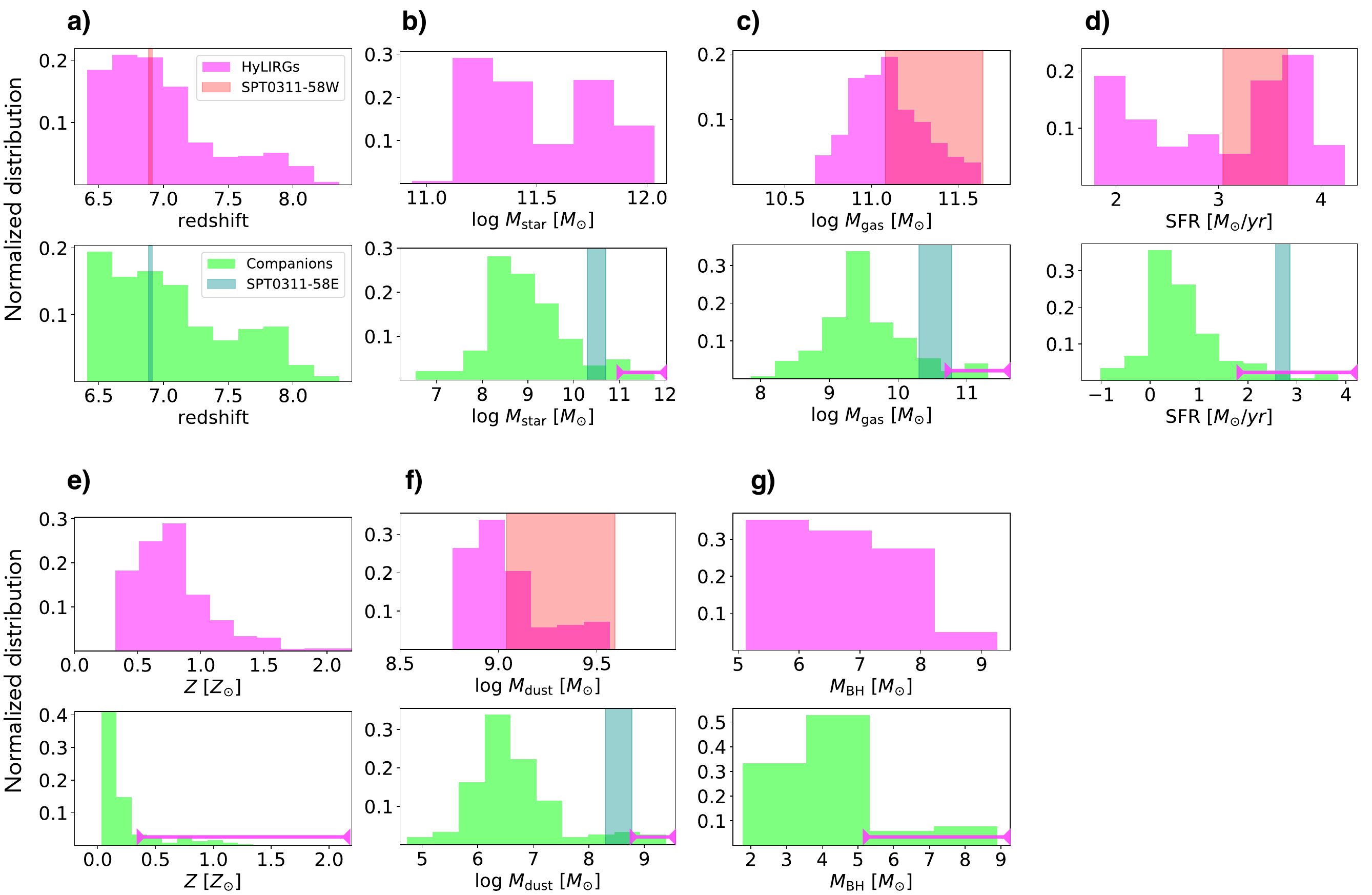}
	\caption{
		%
		The magenta histograms show the distributions of physical properties ($M_{\rm{star}}$, $M_{\rm{gas}}$, $\rm{SFR}$, $Z$, $M_{\rm{dust}}$, $M_{\rm{BH}}$) and redshift of HyLIRGs progenitors  of $z\sim6$ quasars (namely the galaxies included in the magenta filled region of Fig. \ref{fig:L_IR_analysis}).
		%
		The light green histograms show the same distributions for systems experiencing mergers with HyLIRGs progenitors.
		The overlaid magenta lines indicate the range of values of HyLIRGs properties, revealing clear differences between HyLIRGs and their interacting companions.
		The overlaid red and dark green solid areas indicate the observed values of the two galaxies composing the interacting system observed by \protect\cite{Marrone2018}, i.e., SPT0311-58W and SPT0311-58E respectively. 
		The width of the area is representative of the uncertainty on the measurements.
		}
	\label{fig:properties_analysis}
\end{figure*}
\subsection{The physical properties of high-$z$ HyLIRGs}
\label{sec: X}

In Fig. \ref{fig:properties_analysis} we show the redshift, stellar mass ($M_{\rm{star}}$), gas mass ($M_{\rm{gas}}$), SFR, dust mass ($M_{\rm{dust}}$), metallicity ($Z$) and BH mass ($M_{\rm BH}$) distributions of HyLIRG progenitors (magenta histograms) of high-$z$ quasars.
We find that HyLIRG progenitors are generally highly massive (i.e., $M_{\rm{star}} \sim 10^{11}-10^{12} M_\odot$; Fig. \ref{fig:properties_analysis}.b), gas rich (i.e., $M_{\rm{gas}} \sim 10^{10.5}-10^{11.5} M_\odot$; Fig. \ref{fig:properties_analysis}.c) systems. 
They experience strong starbursts, with $\rm{SFRs} \sim 10^2 - 10^4 ~ M_\odot ~ \rm{yr}^{-1}$ (see Fig. \ref{fig:properties_analysis}.d), enabling them to reach super-solar metallicities (see Fig. \ref{fig:properties_analysis}.e) and large dust masses (i.e., $M_{\rm{dust}} \sim 10^{9} M_\odot$; Fig. \ref{fig:properties_analysis}.f) already at $z\sim 6.5 - 7.5$.
The same properties are also shown for galaxies experiencing mergers with HyLIRG companions along the merger trees (light green histograms).
Those galaxies are found to have different properties than HyLIRGs (note the different x-axis range between magenta and green histograms in Fig. \ref{fig:properties_analysis}).
In particular, we find that  the bulk of the population of interacting companions is generally less massive (the $M_{\rm{star}}$ distribution peaks at $M_{\rm{star}} \sim 10^{9} M_\odot$), less star-forming (the SFR distribution peaks at $\rm{SFR} \sim 5 ~ M_\odot ~ \rm{yr}^{-1}$) and less chemically evolved (i.e., sub-solar metallicities and $M_{\rm{dust}} \sim 10^{6.5} M_\odot$).
However, we find that the tails of the distributions of interacting companions extend toward the HyLIRG regime, as shown by the intersections between the light green histograms and the magenta lines in Fig. \ref{fig:properties_analysis}.
The existence of this channel of interaction is crucial from the point of view of evolutionary models of SMBH-galaxy co-evolution: 
in fact, gas-rich major mergers are thought to be an efficient mechanism to deliver cold gas at the centre of the resulting halo and therefore fuel the growth of the SMBH and trigger AGN activity (\citealp{DiMatteo2005,Hopkins2008,Pezzulli2016}).

\begin{figure}
	\centering
	\includegraphics[width=1\columnwidth]{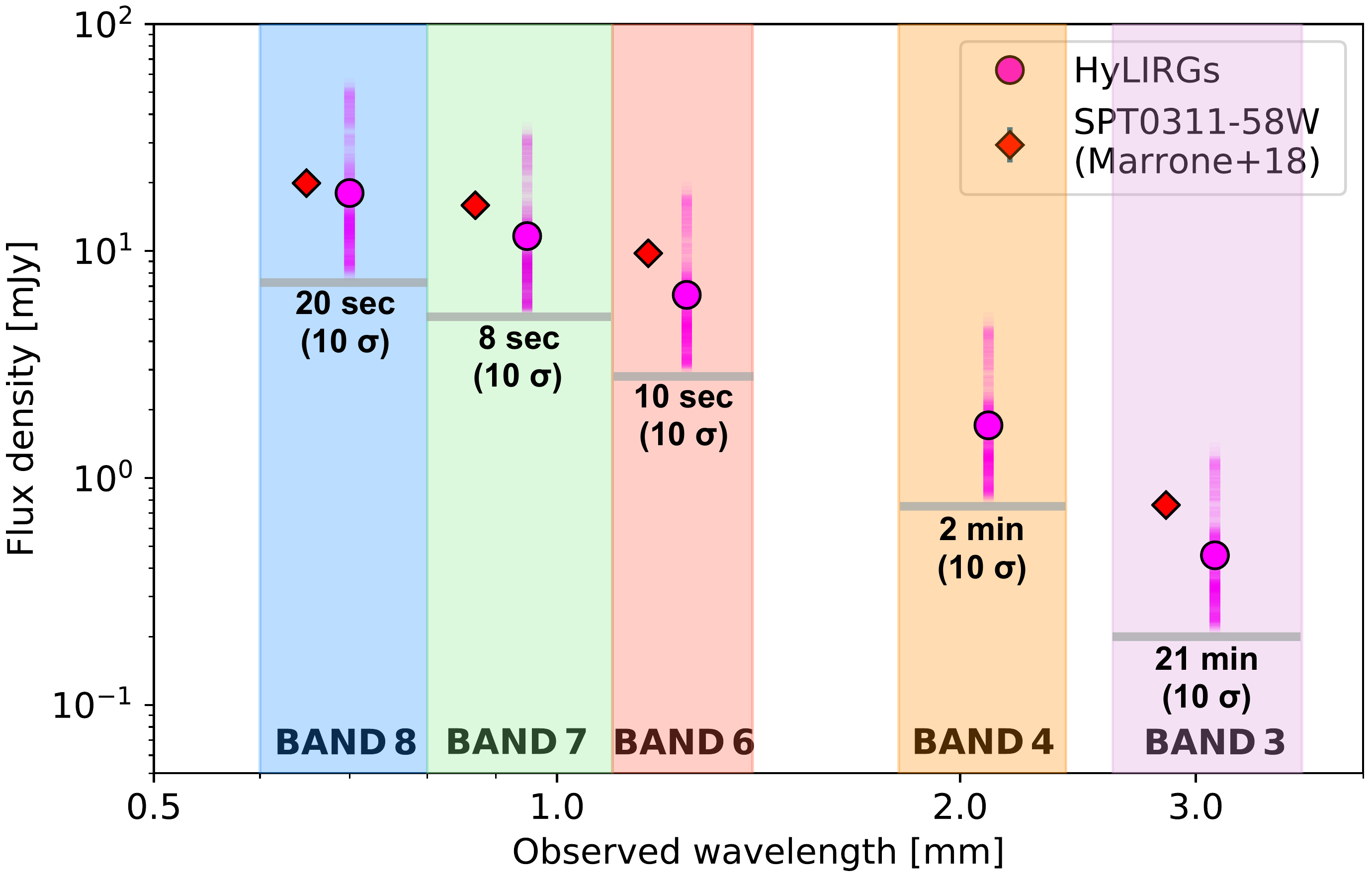}
	\caption{
		Rest-frame far-IR fluxes of HyLIRG progenitors of a J1140-like quasar host galaxy.	Fluxes are calculated in the ALMA Bands [8, 7, 6, 4, 3] (coloured filled regions) respectively centered at observed-frame [0.7, 0.95, 1.25, 2, 3] mm.
		Magenta points represent the averaged values of the HyLIRGs population, and bars represent their dispersions (the colour intensity traces the flux distribution at any given wavelength bin).
		The red points represent the fluxes of SPT0311-58W (\citealp{Marrone2018}) in the ALMA Bands [8, 7, 6, 4, 3].
		For fluxes corresponding to the grey bars, located at the low-end of the HyLIRG  flux distributions at any wavelength bin, we report the on-source observing time needed by ALMA to reach a S/N = $ 10 \, \sigma$.}
	\label{fig:Flux_IR_analysis}
\end{figure}

\subsection{The observability of high-$z$ HyLIRGs with ALMA}\label{sec:ALMA}

By assuming an optically thin, $\tau_{\rm{d}}(\lambda) << 1$, rest-frame far-IR emission, the intrinsic flux observed in a given band $F_{\nu}$ can be estimated as,
\begin{equation}
F_{\nu} = \dfrac{M_{\rm{dust}} ~ k_{\rm{d}}(\nu) ~ B(\nu, T_{\rm{dust}})  ~ (1+z)}{d_{\rm{L}}^2(z)},
\end{equation}
where $d_{\rm{L}}^2(z)$ is the luminosity distance of a source at given redshift.
\newline
However, it is well known that at high redshift the cosmic microwave background (CMB) is hotter and brighter. 
Its temperature, $T_{\rm{CMB}}(z) = T_{\rm{CMB}}^{z=0} ~ (1+z)$, at $z>5$ approaches the temperature of the cold dust, decreasing the contrast of the intrinsic rest-frame IR emission against the CMB.
To account for this effect, we adopted the results by \cite{daCunha2013}, who provided general correction factors to estimate what fraction of the dust emission can be detected against the CMB as a function of frequency, redshift and temperature,
\begin{equation} \label{eq:CMB}
\frac{F_{\nu}^{\rm{against ~ CMB}}}{F_{\nu}^{\rm{intrinsic}}} = 1 - \frac{B(\nu, T_{\rm{CMB}}(z))}{B(\nu, T_{\rm{dust}}(z))}.
\end{equation}
\newline
It follows from Eq. \ref{eq:CMB} that the flux-CMB contrast increases at higher $T_{\rm{dust}}$ (see also the discussion in  \citealt{daCunha2013}).
This is the case for our simulated HyLIRGs, where the interstellar dust in the warm phases of the ISM is heated by the vigorous starburst activity, reaching the high temperature values adopted in this paper ($T_{\rm{dust}}\gtrsim 40 ~\rm{K}$, see also \citealp{Dunne2000, Wang2008, Valiante2016}).
Indeed we find that a large fraction, namely $70 - 80 \%$, of the intrinsic rest-frame IR emission can be detected against the CMB.
\newline
\newline
Fig. \ref{fig:Flux_IR_analysis} shows the distribution (both averaged values and dispersions) of the rest-frame IR fluxes (\textit{CMB-corrected}) for HyLIRG progenitors at $6.4 < z < 9$.
Fluxes are calculated in the ALMA Bands [8, 7, 6, 4, 3], respectively centered at observed-frame wavelengths  [0.7, 0.95, 1.25, 2, 3] mm, and range from a minimum value of $0.2 ~ \rm{mJy}$ at 3 mm (Band 3), to a maximum value of $50 ~ \rm{mJy}$ at 0.7 mm (Band 8). 
We calculated the \textit{on-source} integration times%
\footnote{The \textit{on-source} integration times were obtained using the ALMA Cycle 6 sensitivity calculator.}
needed to detect the estimated continuum fluxes with a signal-to-noise of $\rm{S/N} = 10$.
We find that HyLIRGs progenitors of high-$z$ quasar host galaxies can be easily detected with ALMA with a modest amount of observing time, ranging from few second in Bands 6-7-8, to $\sim20$ minutes in Band 3 (see Fig. \ref{fig:Flux_IR_analysis}).
We note that, although for such short integrations most of the time is dedicated to the instrumental setup and calibrations, the total integration time  (including overheads) needed  to reach the required sensitivity is less than 30 min in all the Bands.

\begin{figure}
	\centering
	\includegraphics[width=1\columnwidth]{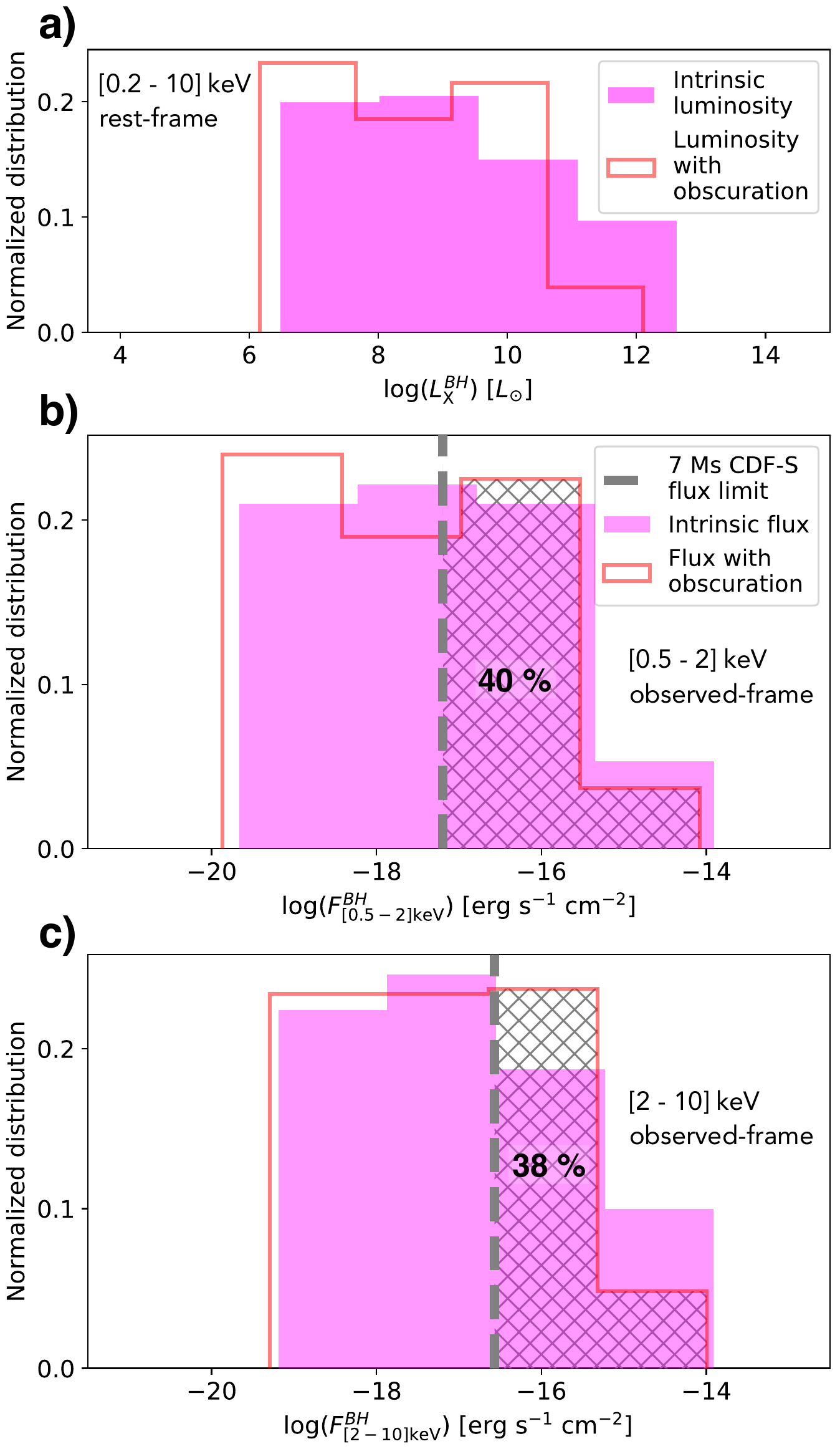}
	\caption{
		\textbf{\textit{a})}
		Distribution of the intrinsic (magenta histogram) and obscured (red histogram) X-ray luminosities (integrated over the rest-frame soft+hard band, [0.2 -10] keV)  produced by the accreting BHs in the HyLIRG progenitors of high-$z$ quasars.
		\textbf{\textit{b})} and \textbf{\textit{c})}
		Distributions of the intrinsic and obscured X-ray fluxes produced by the accreting BHs, in the observed-frame soft (panel \textit{b}) and hard (panel \textit{c}) X-ray bands.
		The grey dashed lines represent the flux limits of the 7 Ms CDF-S survey, i.e., $F_{\rm CDF-S} = 6.4 \times 10^{-18}  {\rm erg ~ s^{-1} ~ cm^{-2}}$ in the soft band, and  $F_{\rm CDF-S} = 2.7 \times 10^{-17}  {\rm erg ~ s^{-1} ~ cm^{-2}}$ in the hard band.
		About 40\% (38\%) of the HyLIRGs population shows a detectable $F_{\rm X}^{\rm BH}$ in the observed-frame soft (hard) band (hatched regions).
	}
	\label{fig:Flux_X}
\end{figure}

\begin{figure}
	\centering
	\includegraphics[width=1\columnwidth]{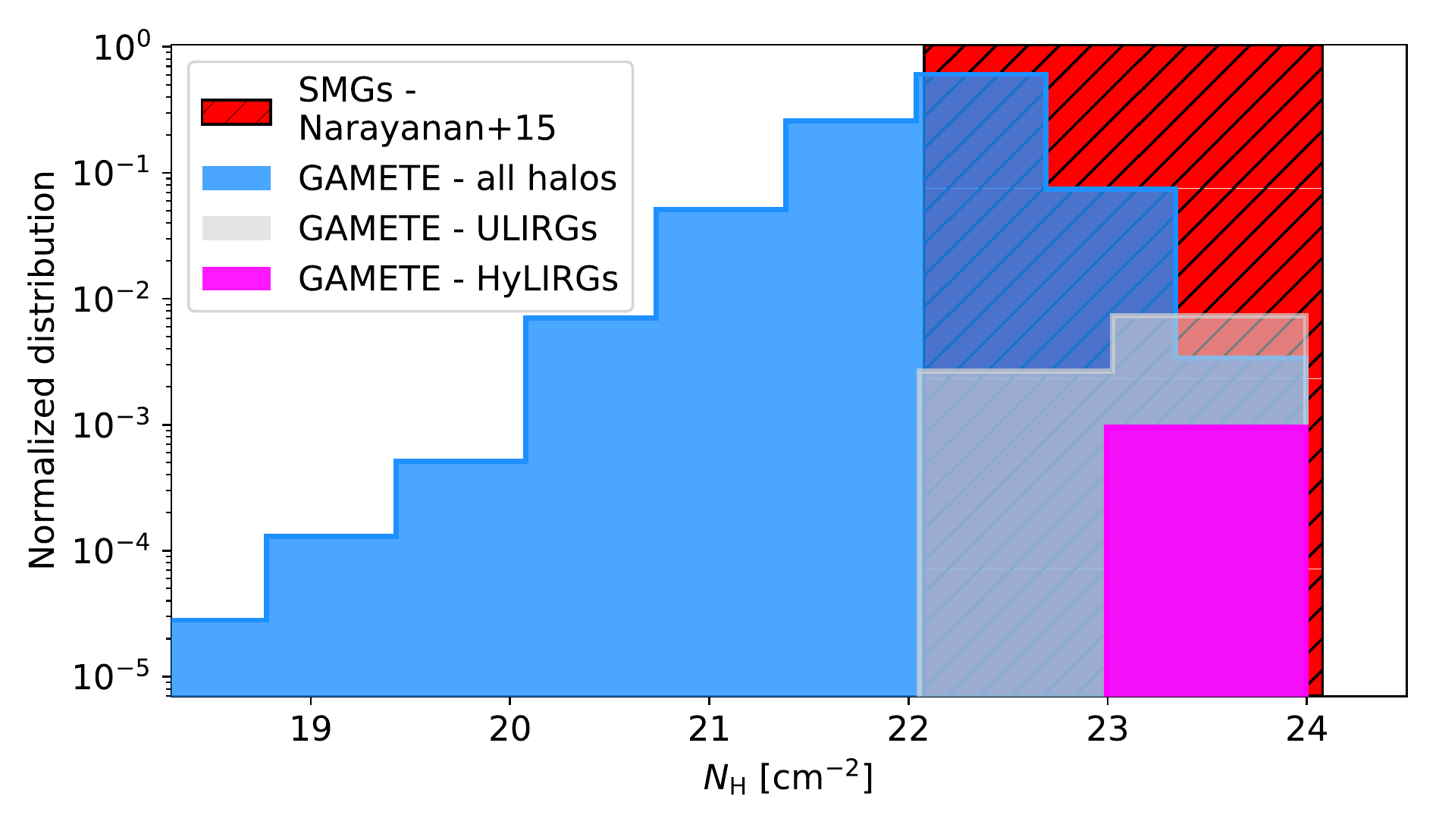}
	\caption{
		The blue histogram shows the distribution of $N_{\rm H}$ predicted for the overall population of galaxies in our model, while the grey and magenta histograms show the distribution of $N_{\rm H}$ predicted for the ULIRG and HyLIRG progenitors of the $z\sim6$ quasar host galaxies, respectively.
		The $N_{\rm H}$ distribution for ULIRGs is consistent with the distribution of $N_{\rm H}$ predicted for SMGs by the cosmological hydrodynamical simulation by \protect\cite{Narayanan2015} (red hatched region).}
	\label{fig:nH}
\end{figure}

\subsection{The X-ray observability of BH activity in high-$z$ HyLIRGs} 

\begin{figure}
	\centering
	\includegraphics[width=1\columnwidth]{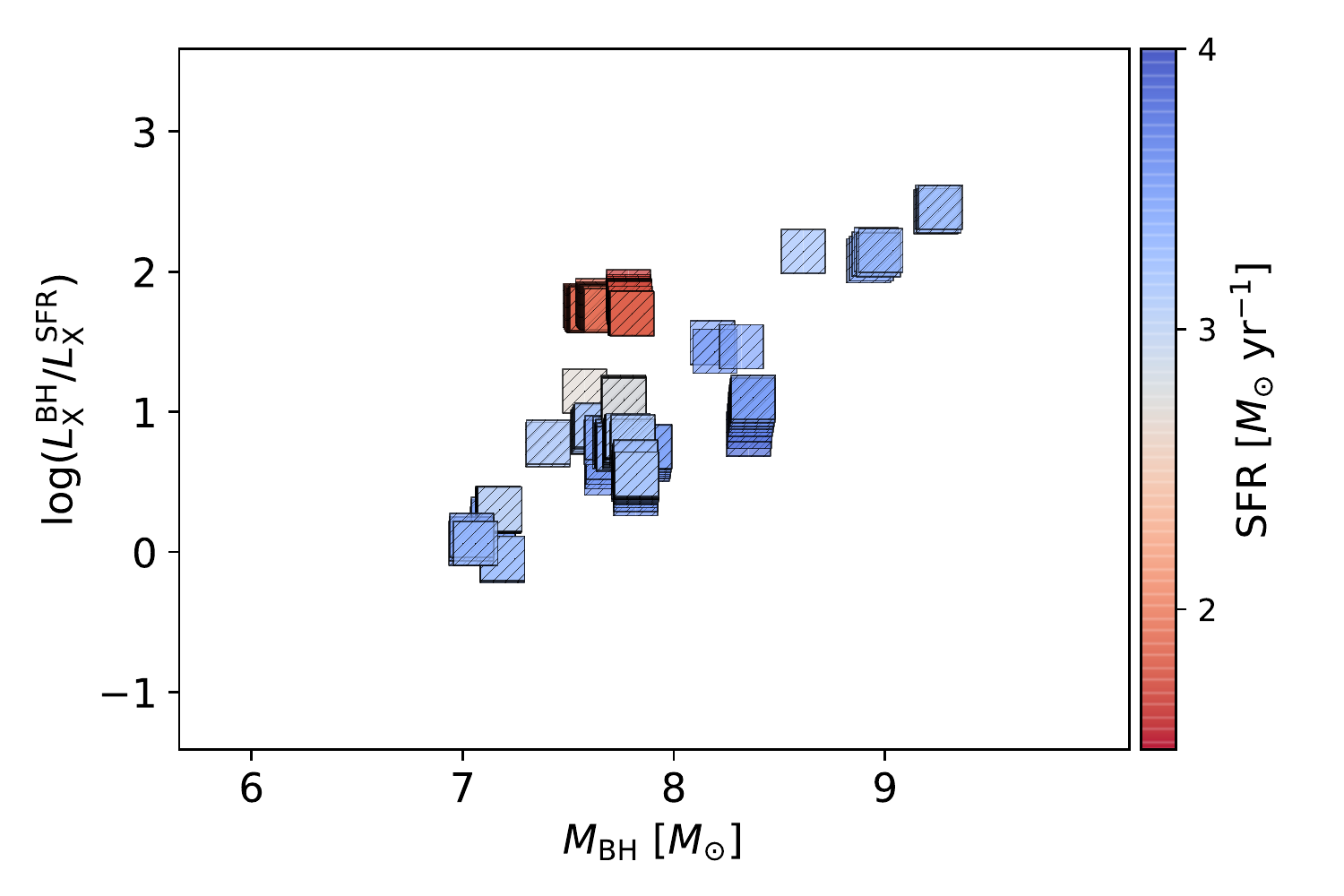}
	\caption{The relative contribution of rest-frame hard X-ray emission produced by nuclear activity and stars is shown as a function of $M_{\rm BH}$ and colour coded depending on the SFR, for galaxies with a detectable observed-frame soft X-ray flux (namely the objects in the red hatched region of Fig. \ref{fig:Flux_X}.b).
	At $M_{\rm BH} \gtrsim 10^7 ~ M_{\odot}$, the X-ray emission produced by accreting SMBHs dominates over the X-ray emission produced by stars.
	For about $40 \%$ of the population,  log($L_X^{\rm BH}/L_X^{\rm SFR}) >1$.
	}
	\label{fig:BH_vs_SFR}
\end{figure}

As shown in Fig. \ref{fig:properties_analysis}.g, we find that the HyLIRG progenitors of luminous high-$z$ quasars, host nuclear BHs with an almost uniform distribution of masses in the range $10^{5} M_{\odot}  < M_{\rm BH}< 10^8 M_{\odot}$, and a high-mass tail at $M_{\rm BH} \lesssim 10^9 M_{\odot}$ (populated by systems in the last steps of the merger trees).
Such a broad range of nuclear BH masses results from the diversity of merging and accretion histories predicted
by the model.
To assess the possibility of detecting BH activity in the nuclei of high-$z$ HyLIRGs, we compared the predicted
X-ray luminosity in the (observed-frame) soft ([0.5 - 2] keV) and hard ([2 - 10] keV) X-ray bands 
with the flux limits of recent surveys.
\newline
Following \cite{Pezzulli2017}, we modeled the X-ray luminosity
of accreting BHs considering the primary emission from the hot corona and 
the reflection component due to the surrounding neutral medium. The first one is
parametrized as a power law,
$L_\nu \propto \nu^{-\Gamma+1} e^{h\nu/E{\rm c}}$, with an exponential cut-off energy 
of $E_{\rm c} = 300\, \rm keV$ \citep{Sozonov2004, Yue2013} and photon spectral index 
$\Gamma = 0.32 \, {\rm log} \,\lambda_{\rm Edd} + 2.27$, that we assume to depend on the Eddington ratio 
$\lambda_{\rm Edd} = L_{\rm bol}/L_{\rm Edd}$ \citep{Brightman2013}. The metallicity
dependent reflection component is computed using the PEXRAV model \citep{Magdziarz1995} 
in the XSPEC code%
\footnote{The reflection component is computed assuming an isotropic source located above the disc, 
fixing the reflection solid angle to $2 \pi$, the inclination angle to $60^{\circ}$ and the reflection strength parameter 
to $R=1$, consistent with typical values of local AGNs \citep{Zappacosta2018}.}. The metallicity and Eddington
ratio $\lambda_{\rm Edd} = L_{\rm bol}/L_{\rm Edd}$ are computed using the values 
predicted by \texttt{GQd} for each accreting BH.
\newline
We find that the BH activity in HyLIRGs generates an intrinsic emission corresponding to X-ray luminosities (integrated over the rest-frame [0.5-10] keV band) of  $L_{\rm X}^{\rm BH} =$ [$10^{6.5}  - 10^{12.5}$] $L_{\odot}$ (see Fig. \ref{fig:Flux_X}.a),
and to intrinsic fluxes $F_{\rm X}^{\rm BH}=$ [$10^{-20}  - 10^{-14}$] $\rm erg ~ s^{-1} ~ cm^{-2}$ in the observed-frame soft/hard bands (see flux distributions in Fig. \ref{fig:Flux_X}.b and Fig. \ref{fig:Flux_X}.c).
\newline
%
The intrinsic X-rays flux is attenuated by the interaction of the radiation produced during the BH accretion with the gas and dust in the immediate surroundings of the BH (mainly photoelectric absorption and Compton scattering of photons against free electrons).
For this reason the effect of obscuration has to be taken into account, and the obscured emerging flux can be written as $F_{\nu}^{\rm abs} = F_{\nu}^{\rm unabs} ~ e^{-\tau_\nu}$; at energies $E \gtrsim 0.1$ keV, assuming a fully-ionized H-He mixture, the optical depth $\tau_{\nu}$ is (\citealp{Yaqoob1997}):
\begin{equation}
\tau_{\nu} = (1.2 ~\sigma_{\rm T} + \sigma_{\rm ph})~ N_{\rm H},
\end{equation}
where $N_{\rm H}$ is the hydrogen column density, while $\sigma_{\rm T}$ and $\sigma_{\rm ph}$ are the Thomson and photoelectric cross section, respectively (see \citealp{Pezzulli2017} for a description of  $\sigma_{\rm ph}$ and its dependence on energy and metallicity).
We computed $N_{\rm H}$ assuming the gas to be distributed following a singular isothermal sphere (SIS) profile (see \citealt{Valiante2016}), 
\begin{equation}
\rho(r)  = \dfrac{\rho_0}{1 + (\frac{r}{r_{\rm core}})^2},
\end{equation}
where the normalization factor $\rho_0$ is a function of $M_{\rm gas}$, the virial radius ($R_{\rm vir}$) and the core radius ($r_{\rm core}$).
\newline
By assuming $r_{\rm core}= 50$ pc, we find that our simulated ULIRGs have $N_{\rm H}  = [10^{22} - 10^{24}] ~ {\rm cm}^{-2}$,  in the same range predicted by cosmological hydrodynamical simulations for high-$z$ SMGs (e.g., \citealp{Narayanan2015}; see Fig. \ref{fig:nH}). 
Adopting the same $r_{\rm core}$, we find that HyLIRG progenitors have $N_{\rm H}  \sim [10^{23} - 10^{24}] ~ {\rm cm}^{-2}$.
%
%
As shown in Fig. \ref{fig:Flux_X}.b and Fig. \ref{fig:Flux_X}.c, the resulting net attenuation of fluxes by obscuration in the observed-frame soft X-ray band is about a factor of 2, while fluxes in the observed-frame hard X-ray band are almost unaffected by attenuation. 
This can be explained in terms of the photoelectric cross section decreasing for increasing energy (see a thorough discussion in \citealp{Pezzulli2017}).
\newline
Finally, we compared the predicted, obscuration-attenuated $F_{\rm X}^{\rm BH}$ with the flux limits obtained by \textit{Chandra} observations of the recent 7 Ms \textit{Chandra} Deep Field South Survey (7 Ms CDF-S; \citealp{Luo2017}), i.e., $F_{\rm CDF-S} = 6.4 \times 10^{-18}  {\rm erg ~ s^{-1} ~ cm^{-2}}$ in the soft band and  $F_{\rm CDF-S} = 2.7 \times 10^{-17}  {\rm erg ~ s^{-1} ~ cm^{-2}}$ in the hard band.
We find that in both bands a significant fraction of our simulated HyLIRGs population has a potentially detectable $F_{\rm X}^{\rm BH}$ (40\% and 38\%, respectively; see hatched regions of the red histograms in Fig. \ref{fig:Flux_X}.b and Fig. \ref{fig:Flux_X}.c).
\newline
To test if $F_{\rm X}$ can be considered as a tracer of nuclear activity in high-$z$ HyLIRGs, we estimated the contribution of SFR-related (e.g., X-ray binary populations; XRBs) processes to the X-ray luminosity, $L_{\rm X}^{\rm SFR}$, by using the redshift-dependent scaling relation between the rest-frame hard $L_{\rm X}$ and SFR, found by \cite{Lehmer2016}:

\begin{equation*}
{\rm log}(L_{\rm X}^{\rm SFR})_{[2-10] ~ {\rm keV}} = 39.82  + 0.63  ~{\rm log(SFR)} + 1.31 ~{\rm log(1+z)}.
\end{equation*}
In Fig. \ref{fig:BH_vs_SFR}, we show the relative contribution between the X-ray luminosity produced by the accreting SMBHs (once the attenuation by obscuration has been considered) and by XRBs, in the rest-frame hard band, as a function of the BH mass (and colour coded with the SFR), for the HyLIRGs progenitors with a detectable observed-frame soft%
\footnote{The observed-frame [0.5-2] keV band traces hard X-ray emission at the redshifts of interest.} $F_{\rm X}^{\rm BH}$ (i.e., namely the objects located in the hatched region of the histograms in Fig. \ref{fig:Flux_X}.b).
We find that HyLIRGs hosting black holes with $M_{\rm BH} \gtrsim 10^{7}~ M_{\odot}$ have X-ray emission dominated by their SMBH activity. 
In particular, we find that, for about 40 \% of the these galaxies, the X-ray luminosity due to the nuclear activity is at least one order of magnitude larger than that
associated to star formation. Since we are conservatively assuming that the X-ray emission produced by stars is completely unobscured, if $L_{\rm X}^{\rm SFR}$ were 
corrected for attenuation%
\footnote{\texttt{GQd} model cannot provide information on the spatial distribution of stars, therefore a self-consistent treatment of 
obscuration of the stellar X-ray emission is very difficult (see \citealt{Valiante2016}), without coupling detailed hydrodynamical models with radiative transfer simulations accounting for X-ray physics (\citealp{Smidt2017, Graziani2018}). 
On the other hand a statistical approach is currently impossible due to computational limits.}, the relative contribution of SMBH activity to the total X-ray emission 
would be even higher. 
\newline
In conclusions, we find that (i) a significant fraction of  HyLIRG progenitors has detectable (obscuration-attenuated) X-ray fluxes produced by accreting BHs in both the observed-frame soft (40 \%) and hard (38 \%) bands; (ii) for those objects, the overall $L_{\rm X} = L_{\rm X}^{\rm SFR} + L_{\rm X}^{\rm BH}$ is dominated by $L_{\rm X}^{\rm BH}$, making the observed X-ray emission an optimal tracer of SMBH activity.
Combining these findings with the predicted spatial density of HyLIRGs (see Sec. \ref{sec: spatial_density}), we estimate the spatial density of HyLIRGs with detectable 
X-ray emission tracing SMBH activity to be  $\sim 0.07~ {\rm Gpc^{-3}}$.

\section{Conclusions} \label{sec: conclusions}

In the last decade, observations of DSFGs have extended toward $z>5$.
High-$z$ DSFGs are often found to be HyLIRGs ($L_{\rm{IR}} > 10^{13} L\odot$) with SFRs exceeding $1000 M_\odot \rm{yr}^{-1}$, likely tracing the result of strong dynamical interactions and intense gas accretion events occurring in the densest regions of the early Universe
(\citealp{Walter2012, Riechers2017, Pavesi2018}).
Recently, \cite{Marrone2018} found a IR-luminous system, SPT0311-58, at $z \sim 7$, composed by a pair of extremely massive and IR-luminous interacting galaxies.
Using different proxies to estimate the halo mass hosting the system, they show that SPS0311-58 marks an exceptional peak in the cosmic density field at this early cosmic time and it lies close to the exclusion curve predicted by the current structure formation paradigm. 
\newline
In this work we use \texttt{GQd} model to simulate several independent merger histories of a typical luminous quasar at $z\sim 6$, following the black hole growth and the baryonic evolution (including star formation, metals and dust life-cycles, and mechanical feedback) of its ancestors up to $z\sim 10$.
We find that:
\begin{itemize}
\item[-] a fraction of progenitor galaxies (about 0.4 objects per single luminous quasar, between $z\sim6-8$) has HyLIRGs-like IR luminosity of $L_{\rm{IR}}>10^{13} ~L_{\odot}$ (see Fig. \ref{fig:L_IR_analysis}) and similar characteristics to the system observed by \cite{Marrone2018} (e.g., generally highly massive, extremely star-bursting and chemically evolved; see Fig. \ref{fig:properties_analysis});
\newline
\item[-] the IR-luminous progenitors of $z\sim6$ quasar host galaxies, reside in the most massive DM halos, with masses of $M_{\rm{DM}} \sim 10^{12.5}-10^{13} ~ M_\odot$ (see Fig. \ref{fig:DM_analysis}).
Their far-IR continuum fluxes can be easily observed with ALMA with a modest amount of time, ranging from few seconds in Bands [8,7,6] (centered at observed-frame [0.7, 0.95, 1.25] mm, respectively) to about 2 minutes in Band 4 (2 mm) and 20 minutes in Band 3 (3 mm).
\newline
\item[-] the HyLIRGs progenitors of high-$z$ quasar host galaxies, have nuclear black holes with masses $10^{5} M_{\odot}  < M_{\rm BH}< 10^8 M_{\odot}$. 
The X-ray fluxes associated to the nuclear BH activity could be detectable in the observed-frame soft and hard X-ray bands (they lie above the sensitivity limit of
the 7 Ms {\it Chandra} Deep Field South Survey) for $\sim 40 \%$ of these galaxies, even accounting for obscuration, and dominate over the X-ray emission
associated to star formation. 
\end{itemize}	 
Altogether our results suggest that $z \sim 6$ luminous quasars are indeed the signposts of the observed rare high-$z$ overdensities, and that massive-IR luminous galaxies at higher $z$ are their natural ancestors. These findings corroborate models of SMBH-galaxy co-evolution predicting an evolutionary link between IR-luminous galaxies and quasars.

\section*{Acknowledgements}

The authors would like to thank D. Marrone for helpful discussions.
The research leading to these results has received funding from the European Research Council under the European Union's Seventh Framework Programme (FP/2007- 2013)/ERC Grant Agreement n. 306476.
We have benefited from the public available programming language \texttt{Python}, including the \texttt{numpy}, \texttt{matplotlib} and \texttt{scipy}  packages.
This research made extensive use of \texttt{ASTROPY}, a community-developed core \texttt{Python} package for Astronomy (\citealp{Astropy2013}), and \texttt{glueviz}, a \texttt{Python} library for multidimensional data exploration (\citealp{Beaumont2015}).



\bibliographystyle{mnras}
\bibliography{biblio.bib} 




\bsp	
\label{lastpage}
\end{document}